# Characterization of IRS-aided Indoor Wireless Virtual-Reality with Hybrid Beamforming

Nasim Alikhani, Abbas Mohammadi,
AUT-Wireless Research Lab.
Electrical Engineering Department,
Amirkabir University of Technology
Tehran, Iran

abm125@aut.ac.ir

*Abstract*— **This paper introduces an optimum solution for a utility function that increases spectral efficiency in wireless Virtual Reality (VR) systems. This system uses Multi-user Multiple Input Multiple Output Orthogonal Frequency Division Multiplexing (MU-MIMO OFDM) with hybrid beamforming in indoor Intelligent Reflecting Surface (IRS) based Downlink (DL) scenario. Given the critical need to maximize the rate for transmitting VR traffic to meet the low-latency requirements, a substantial bandwidth allocation is essential. This bandwidth is assumed to be in the mmWave band, according to the IEEE 802.11ad/ay standard. The proposed utility function takes into account various delays, including processing, transmission and queuing delays, on both DL and Uplink (UL). Moreover, the relation between transmission delay and the utility function is examined in different Signal-to-Noise Ratio (SNR) levels, using both mean and minimum channel gain metrics. An optimization approach is applied to iteratively determine the IRS phase shifts and effective channel gain. The simulation results are benchmarked against NS3 simulations, showing a high degree of consistency. With an average accuracy of 81.57% the calculated DL and UL rates match the NS3 results when considering the IRS. Also, our proposed method achieves superior performance in the case of complexity over the existing designs.**

*Index Terms*— **Virtual Reality (VR), QoS, mmWave, IEEE802.11ay, MU-MIMO-OFDM, utility function, hybrid beamforming, IRS.**

## I. INTRODUCTION

Immersive Virtual Reality (VR) experiences have attracted significant interest from the industry, particularly in terms of delivering high-definition content [1-5]. However, several challenges persist in wireless networks delivering VR, such as ensuring high-resolution VR content, accommodating user mobility and minimizing transmission latency [6]. Existing Wi-Fi standards in Wireless Local Area Networks (WLANs) are inadequate for meeting the required Quality of Service (QoS) levels, particularly in environments with high user density [7]. On the other hand, mmWave technology offers substantial advantages, including high data transmission rates and increased bandwidth. The IEEE 802.11ay standard, which considers the mmWave frequency range, delivers extremely high throughput—up to 100 Gbps in the Downlink (DL)—along with ultra-low latency performance. In the Multi-user (MU)- MIMO framework, multiple users can receive simultaneous service from a single transmitter, such as an Access Point (AP) [8]. Hybrid beamforming is employed in mmWave systems, integrating both digital and analog components. Low-cost and energy-efficient Intelligent Reflecting Surface (IRS) can play a key role in mmWave systems by establishing a direct communication link between the transmitter and the user, even in the presence of obstacles [9]. Key factors that influence VR services include tracking accuracy, transmission delays, processing delays and queuing delays [10]. Ensuring high data rates in VR applications remains a significant challenge due to processing and transmission delays [10]. Recent studies highlight the potential of integrating IRSs into conventional multiuser wireless networks to address these issues [11-13]. Jalali et al. explored IRS configurations aimed at enhancing energy efficiency and optimizing admission control for IoT devices operating with short packets [11]. Additionally, Zhou et al. tackled latency reduction in a secure, multi-user IRS-enabled VR communication setup under conditions of imperfect CSI [13]. However, based on our current understanding, no existing research has focused on optimizing an IRS-powered indoor VR network within a constrained 3D environment while considering its impact on processing, transmission and queuing delays. Furthermore, comparisons of this metric across different codebook strategies remain unexplored. IRS have been introduced to enhance mmWave systems, with much of the research emphasizing sum rate maximization while adhering to power and unit modulus constraints [9, 14].

This paper proposes a novel framework for optimizing a novel multi-attribute utility function-constrained hybrid analog and digital beamforming strategy in mmWave systems. Additionally, IRS is integrated to enhance mmWave communications in VR applications. We propose a multi-attribute utility function that integrates all these factors, based on computed data rates in both DL and Uplink (UL) conditions. This innovation is not considered in VR papers [1-5, 10, 15, 16] . To optimize the phase shifts of IRS elements, we apply the Riemannian Conjugate Gradient (RCG) method while separately optimizing the hybrid beamforming matrices. In this paper, a novel multi-attribute utility function is proposed that is based on three types of delay: transmission, processing and queuing delay.The proposed hybrid beamforming method exhibits low computational complexity in the calculation of hybrid beamforming matrices for Multi-user Multiple Input Multiple Output Orthogonal frequency-division multiplexing (MU-MIMO OFDM) systems. The considered scenarios include the mean and minimum channel gain cases. A nonconvex optimization problem based on a proposed multiattribute utility function is formulated. To solve this function with coupled variables, an efficient algorithm based on Alternating Optimization (AO) is developed. Simulation results for an IRS-aided system based on IEEE 802.11ay in an indoor environment using NS3 and MATLAB are provided. The results of MATLAB and NS3 simulations are compared using six different codebooks generated in NS3: "2 Antennas, 1 RF chain", "2 Antennas, 2 RF chains", "4 Antennas, 1 RF chain", "4 Antennas, 2 RF chains", "8 Antennas, 1 RF chain" and "8 Antennas, 2 RF chains".

The rest of the paper is organized as follows: In Section II, system model is presented. In Section III, delay considerations and the utility function are described. In Section IV, the details of the solution approach adopted for addressing the optimization problem are outlined. In Section V, the indoor area and simulation results are explained. Finally, the conclusion is presented in Section VI.

Notation: A matrix is represented by a boldface alphabetic character such as x, a vector is represented in italic format with an underline such as $\underline{x}$, and scalar numbers are represented by $x$. $(.)^{-1}$ and $(.)^H$ denote the inverse and hermitian of a complex matrix, respectively. ∘ stands for the Hadamard (elementwise) product.

## II. SYSTEM MODEL

In this section, we first present the optimization problem. We then describe the channel models for both the UL and DL, along with the relevant delay factors and the formulation of the utility function in the presence of IRS elements.

As depicted in Fig. 1, the system consists of two APs and four users. Because obstacles may weaken the Line-of-Sight (LoS) path, the IRS is introduced to improve the connection between the users and the APs. For this reason, IRS elements are placed between the APs and the users. The solid black lines in Fig. 1 indicate the main DL beam directions — from the AP to the IRS and from the IRS to the user. The dashed black lines show the main UL beam directions — from the user to the IRS and then from the IRS to the AP. Each AP can be equipped with 1, 2, or 4 antennas, while each user has a single antenna. However, the formulations of DL and UL channels are illustrated based on MU-MIMO OFDM.

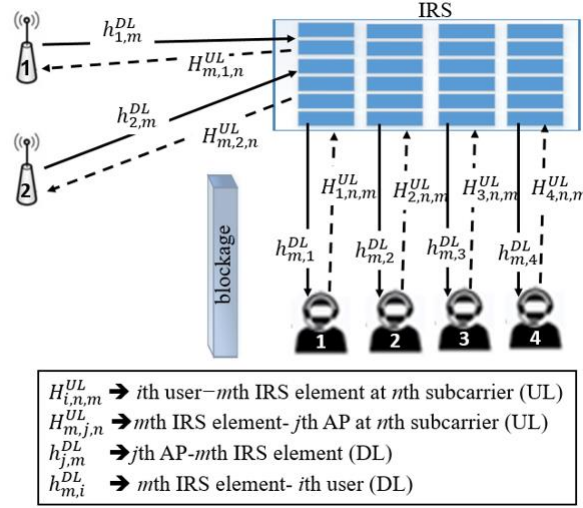

**Fig. 1.** System Model configuration.

### A. Optimization problem

The goal of the proposed method is to simultaneously determine the maximum QoS requirements for users in VR systems. The corresponding optimization problem is formulated in (1). This problem can also be expressed as finding the optimal closed-form solution for the utility function. Since the problem is non-convex, its formulation is presented in (1).

$$\max_{\underline{U}_j, w_i^n, f_i^n, \boldsymbol{\phi}} \sum_{j \in B} \sum_{i \in \underline{U}_j} \sum_{n=1}^{N_{sc}} \mathbf{U_i}(\mathbf{D}_{ijn}, K_{ijn}) \tag{1}$$

$$s.t. |\underline{U}_j| \leq \underline{V}_j, \quad \forall j \in B \tag{a}$$

$$c_{ij}^{DL} \geq R_{ij}^{min}, \quad \forall j \in B, \forall i \in \underline{U}_j \tag{b}$$

$$\mathbf{P}_{A_i} \in \underline{N}_{AP}^j, \quad [\mathbf{P}_{A_i}]_{k,l}[\mathbf{P}_{A_i}]_{k,l}^* = \frac{1}{N_t} \tag{c}$$

$$\mathbf{G}_{A_i} \in \underline{N}_{STA}^j, \quad [\mathbf{G}_{A_i}]_{k,l}[\mathbf{G}_{A_i}]_{k,l}^* = \frac{1}{N_r} \tag{d}$$

$$|e^{j\phi_{m,m}}| = 1, \quad \forall \phi_{m,m} \in \boldsymbol{\phi} \tag{e}$$

3$U_i(D_{ijn}, K_{ijn})$ is the equivalent utility function between $i$th user, $j$th AP at $n$th subcarrier. This function is expressed thoroughly in section III. Constraint (a) is related to the number of users connected to the $j$th AP. The second constraint (b) denoted that the minimum level of SINR in DL for the $i$th user in contact with the $j$th AP is at least equal to $R_{ij}^{min}$. $F^{in} = P_{A_i}^H P_{digital(i,n)}$ represent the precoder matrix and $W^{in} = G_{A_i} G_{digital(i,n)}$ represents the combiner matrix. Both $P_{A_i}$ and $G_{A_i}$ have unit amplitude as denoted in (c) and (d) respectively. $[P_{A_i}]_{k,l}$ is the element in the $k$th row and $l$th column of the $P_{A_i}$ matrix in (c). $\underline{N}_{AP}^j$ is the set of codebooks in the $j$th AP and $[G_{A_i}]_{k,l}$ is the element in the kth row and lth column of the $G_{A_i}$ matrix in (d). $\underline{N}_{STA}^j$ is the number of codebooks for the $i$th user. In this paper we consider one type of codebook for user that it contains one antenna and one RF chain. The number of RF chain for each AP is denoted as $N_{RF}$. $\boldsymbol{\phi}$ is the total matrix of phase shift elements of IRS as denoted in (e). $\phi_{m,m}$ is equal to the mth phase shift element in the main diagonal values of matrix $\boldsymbol{\phi}$. $U$ is the total number of users. $B$ is the total number of APs.

*B. UL channel*

The UL channel operates within the sub-6 GHz frequency range [10].

The UL channel matrix consists of three components: $\mathbf{H}_{i,n,m}^{UL}$, $\mathbf{H}_{m,j,n}^{UL}$ and $\mathbf{h}_{ij}^{NLoS}$. Here $N_t, N_r, N_{SC}$ represent the number of antennas at each AP, the number of antennas at each user, and the total number of subcarriers, respectively. The matrix $\mathbf{H}_{i,n,m}^{UL}$ describes the LoS channel between the $i$th user and the $m$th element of IRS on the $n$th subcarrier, with dimensions $N_{sc} \times N_t \times N_r$. Similarly, $\mathbf{H}_{m,j,n}^{UL}$ represents the LoS channel path between the $m$th element of IRS and the $j$th AP at $n$th subcarrier, with the same dimensions. Finally, the channel $\mathbf{h}_{ij}^{NLoS}$ denotes the NLoS channel between the $i$th user and the $j$th AP on the $n$th subcarrier, also with dimensions $N_{sc} \times N_t \times N_r$. If the total number of IRS elements is $N_{IRS}$, the complete UL channel can be expressed as (2) [10]:

$$\mathbf{H}_{ijn}^{UL} = \mathbf{h}_{ij}^{NLoS} + \sum_{m=1}^{N_{IRS}} \mathbf{H}_{i,n,m}^{UL} \phi_{m,m} \mathbf{H}_{m,j,n}^{UL} \qquad (2)$$

$\boldsymbol{\phi}$ represents the overall phase shift matrix of the IRS, where each diagonal element of the matrix $\boldsymbol{\phi} \in \mathbb{C}^{N_{IRS} \times N_{IRS}}$ is denoted by $\phi_{m,m}$. The UL channel matrix $\mathbf{H}_{ijn}^{UL}$ is defined as the sum of the NLoS channel between the AP and the user, and the cascaded user–IRS–AP channel paths. This combined channel has dimensions of $N_{sc} \times N_t \times N_r$.

*C. DL channel*

In mmWave technology, the DL channel calculations take into account the number of required bit streams per user, denoted as Ns, and the number of RF chains in the antenna configuration of each AP, denoted as $N_{RF}$.

*1) Calculation of the parameters in DL channel*

In the DL, two channels are considered: 1) The channel between the user and the IRS elements and 2) The channel between the AP and the IRS elements. Each of these channels has its own separate channel gain.

By considering the position of the AP as the transmitter (TX) and the position of the IRS elements as the receiver (RX) in the channel between the IRS and each AP in the DL, the values of the Direction of Departure (DOD) and Direction of Arrival (DOA) can be computed. Specifically, the DOD is given by DOD=RX-TX and the DOA is calculated as DOA=- DOD for the channel between the AP and the IRS. Similarly, the same operations are applied to the channel between the IRS and the user, where the TX represents the location of the IRS elements and the RX represents the location of the user. If we consider that the elements of the IRS are positioned in the Y-Z plane, and the number of elements in the Y and Z directions are $M_z$ and $M_y$, respectively, then the total number of elements in the IRS is given by $M = M_y \times M_z$.

By using the values of DOA and DOD, the Angle of Arrival (AOA) and Angle of Departure (AOD) in both the azimuth and elevation directions can be calculated. A Uniform Linear Array (ULA) antenna configuration [9] is considered for both the user and the AP sides. The steering vectors for this antenna type are denoted as $\underline{AOD}_{AZ}^v, \underline{AOA}_{AZ}^v$. A Uniform Planar Array (UPA) antenna [9] configuration is considered for the IRS. The steering vector for this antenna type in the channel between the AP and the mth element of the IRS is denoted as $\underline{AOA}_m^v$. The format of the DL channel is adapted from [16]. gain in the channel between the IRS and the AP is denoted as $\mathbf{pg}_{mjn}$ (between the $m$th IRS element and the $j$th AP in the $n$th subcarrier). $w$ is used as path-loss exponent for computing this channel gain. Therefore, the complex channel in the DL between the AP and the $m$th element of the IRS is as follows:

$$\mathbf{h}_{j,m}^{DL} = \sum_t \sum_{n=1}^{N_{SC}} \left( 10^{\frac{\mathbf{pg}_{mjn}}{10}} \times \underline{AOD}_{AZ}^v \times \underline{AOA}_m^v \times e^{-\frac{t}{\tau_m}} \right) \qquad (3)$$

Where $\underline{\tau_m} = \frac{\mathbf{d}_{m,j}}{c}$ is the estimated delay between the $m$th element and the $j$th AP, with $\mathbf{d}_{m,j}$ being the distance between the $m$th

IRS element and the *j*th AP, and c being the speed of light. In $\mathbf{h}_{j,m}^{DL}$, ther term $\underline{AOD_{AZ}^v}$ is the steering vector for the AP, and $\underline{AOA_m^v}$ is the steering vector for the *m*th element of the IRS. $\mathbf{pg}_{imn}$ is the channel gain between the *i*th user and the *m*th IRS element in the *n*th subcarrier. The complex channel in the DL between the *m*th element of the IRS and the *i*th user is as follows:

$$\mathbf{h}_{m,i}^{DL} = \sum_t \sum_{n=1}^{N_{SC}} (10^{\frac{\mathbf{pg}_{imn}}{10}} \times \underline{AOD_m^v} \times \underline{AOA_{AZ}^v} \times e^{-\frac{t}{\underline{\tau_m}}}) \tag{4}$$

Where $\underline{\tau_m} = \frac{\mathbf{d}_{i,m}}{c}$ is the estimated delay between the user and the *m*th element of the IRS, with $\mathbf{d}_{i,m}$ being the distance between the *i*th user and the *m*th element of the IRS, and *c* being the speed of light. In $\mathbf{h}_{m,i}^{DL}$, the term $\underline{AOA_{AZ}^v}$ is the steering vector for the user, and $\underline{AOD_m^v}$ is the steering vector for the *m*th element of the IRS. The complex channel gain in the NLoS path between each user and the AP is denoted as $\mathbf{h}_{ij}^{DL}$. The variable *t* represents the total time duration for processing the DL channel computation.

$$\mathbf{h}_{ij}^{DL} = \sum_t \sum_{n=1}^{N_{SC}} (10^{\frac{\mathbf{pg}_{ijn}}{10}} \times \underline{AOD_{AZ}^v} \times \underline{AOA_{AZ}^v} \times e^{-\frac{t}{\underline{\tau}}}) \tag{5}$$

Where $\mathbf{pg}_{ijn}$ is calculated based on the distance between the *i*th user and the *j*th AP in the *n*th subcarrier. $\underline{\tau}$ is the total time vector for DL rate calculation. Thus, the total channel in the DL is:

$$\mathbf{H}_{ij}^{DL} = \mathbf{h}_{ij}^{DL} + \sum_{m=1}^{N_{IRS}} \mathbf{h}_{j,m}^{DL} \phi_{m,m} \mathbf{h}_{m,i}^{DL} \tag{6}$$

$\boldsymbol{\phi}$ is the matrix of complex values of reflection coefficients for the IRS, where each element of the main diagonal values of this matrix is denoted as $\phi_{m,m}$ representing the reflection coefficient of *m*th element of IRS.

*2) Beamforming*

Beamforming techniques are employed in mmWave systems to minimize hardware costs. The beamforming method is used only in the DL direction. Analog beamforming matrices in both the AP-IRS and IRS-user channels are considered constant at each subcarrier frequency.

*a) Hybrid beamforming for calculating the digital beamforming matrix*

In the hybrid beamforming approach, each antenna is connected to a series of phase shifters to reduce energy consumption, which is particularly important in mmWave frequency bands due to high hardware complexity and power requirements. Therefore, the channel matrix must first be updated by incorporating the analog beamforming matrix.

The DL channel matrix after applying analog beamforming is given by: $\mathbf{H}_{D_{SC}} = \mathbf{G}_A^H \mathbf{H}_{SC,AP-IRS} \mathbf{P}_A$. The baseband channel matrix at each subcarrier is denoted as $\mathbf{H}_{eff_{SC}}$ and is calculated as: $\mathbf{H}_{eff_{SC}} = \mathbf{G}_{D_{SC}}^H \mathbf{H}_{D_{SC}} \mathbf{P}_{D_{SC}}$.

To compute the digital combiner $\mathbf{G}_{D_{SC}}^H$, and the digital precoder $\mathbf{P}_{D_{SC}}$ at each subcarrier, the SVD of the updated channel $\mathbf{H}_{D_{SC}}$ is performed as $\mathbf{H}_{D_{SC}} = \mathbf{U}_{D_{SC}} \mathbf{W}_{D_{SC}} \mathbf{V}_{D_{SC}}^H$.

The digital beamforming matrices at each subcarrier frequency matrices are derived from the SVD of the updated channel and are used to optimize the signal transmission and reception in the baseband domain for both the AP-IRS and IRS-user channels, they are expressed as follows:

$$\widehat{\mathbf{P}}_{D_{SC}} = \mathbf{V}_{D_{SC}} \times [I_{N_{SC}} \quad 0_{N_{DS} \times (N_{RF} - N_{DS})}] \tag{7}$$

$$\widehat{\mathbf{G}}_{D_{SC}} = \mathbf{U}_{D_{SC}} \times [I_{N_{SC}} \quad 0_{N_{DS} \times (1 - N_{DS})}] \tag{8}$$

The number of RF chains on the user side is set to 1. Therefore, the following equalities can be concluded as $\widehat{\mathbf{G}}_{D_{SC}} = \widehat{\mathbf{G}}_{D_{SC(i,n)}}$ and $\widehat{\mathbf{P}}_{D_{SC}} = \widehat{\mathbf{P}}_{D_{SC(i,n)}}$. This indicates that the digital combiner and digital precoder matrices are considered for each user *i* and subcarrier *n*. Since the digital precoder depends on the transmitter's power, it is appropriate to normalize the digital precoder by the power of the AP in the DL.

Now, the equations for the digital beamformer in the IRSuser channel are evaluated as follows:

$$\mathbf{H}_{D_{SC}} = \mathbf{G}_A^H \mathbf{H}_{SC,AP-IRS} \mathbf{P}_A \tag{9}$$

$$\mathbf{H}_{D_{SC}} = \mathbf{U}_{D_{SC}} \mathbf{W}_{D_{SC}} \mathbf{V}_{D_{SC}}^H \tag{10}$$

$$\widehat{\mathbf{P}}_{D_{SC}} = \mathbf{V}_{D_{SC}} \times [I_{N_{SC}} \quad 0_{N_{DS} \times (N_{RF} - N_{DS})}] \tag{11}$$

$$\widehat{\mathbf{G}}_{D_{SC}} = \mathbf{U}_{D_{SC}} \times [I_{N_{SC}} \quad 0_{N_{DS}\times(1-N_{DS})}] \tag{12}$$

For updating the channel values (both AP-IRS and IRSuser channels) in the DL based on the calculated hybrid beamforming matrices, the total precoder and total combiner matrices are formed by combining the analog and digital beamforming matrices.

Specifically, for the AP-IRS channel, these matrices are given in (13). These combined matrices are then used to update the channel values and optimize signal transmission and reception in the DL direction.

$$\mathbf{F_1}_{N_t \times N_s} = \mathbf{P}_A \widehat{\mathbf{P}}_{D_{SC}}, \mathbf{W_1}_{N_t \times N_s} = \widehat{\mathbf{G}}_{D_{SC}} \mathbf{G}_A \tag{13}$$

Thus, the updated channel between the IRS elements and the AP in the DL is given by $\mathbf{h}_{j,m,new} = \mathbf{W_1} \mathbf{h}_{j,m}^{DL} \mathbf{F_1}$. Where $\mathbf{W_1}$ and $\mathbf{F_1}$ represent the total combiner and total precoder matrices, respectively, for the AP-IRS channel. Similarly, the channel between the IRS and the user can be updated by the same method. For computing the SINR in the DL, both intra-cell and intercell interference must be taken into account. Intra interference refers to interference from other users within the same cell, while Inter interference refers to interference coming from neighboring cells. The SINR equations for both the UL and DL are presented in (14) and (15) at the beginning of the next page.

Where $P_{u_i}$ represents the transmit power of the $i$th user. $\underline{U}_b$ is the set of users within the coverage area of the $b$th AP. $B$ is the total number of APs. $P_{B_j}$ is the transmit power of the $j$th AP. $\sigma^2$ denotes the power of the complex Gaussian noise, which is independent of both inter- and intra-cell interference.

$$\mathbf{SINR}_{ijn}^{UL} = \frac{P_{u_i}\left(\left(\mathbf{h}_{ijn}^{NLoS}\right)^2 + \sum_{m=1}^{N_{IRS}}\left(\mathbf{H}_{i,n,m}^{UL}\phi_{m,m}\mathbf{H}_{m,j,n}^{UL}\right)^2\right)}{\sigma^2 + \sum_{l\neq i} P_{u_l}\left(\left(\mathbf{h}_{ljn}^{NLoS}\right)^2 + \sum_{m=1}^{N_{IRS}}\left(\mathbf{H}_{l,n,m}^{UL}\phi_{m,m}\mathbf{H}_{m,j,n}^{UL}\right)^2\right) + \mathbf{A}} \tag{14}$$

$$\mathbf{A} = \sum_{b\neq j}^{B}\sum_{l'\neq i}^{U_b} P_{u'_l}\left(\left(\mathbf{h}_{l'bn}^{NLoS}\right)^2 + \sum_{m=1}^{N_{IRS}}\left(\mathbf{H}_{l',n,m}^{UL}\phi_{m,m}\mathbf{H}_{m,b,n}^{UL}\right)^2\right)$$

$$\mathbf{SINR}_{ij}^{DL} = \frac{P_{B_j}\left(\left(\mathbf{h}_{ij}^{DL}\right)^2 + \sum_{m=1}^{N_{IRS}}\left(\mathbf{h}_{j,m,new}^{DL}\phi_{m,m}\mathbf{h}_{m,i,new}^{DL}\right)^2\right)}{\sigma^2 + \sum_{l\neq i} P_{B_j}\left(\left(\mathbf{h}_{lj}^{DL}\right)^2 + \sum_{m=1}^{N_{IRS}}\left(\mathbf{h}_{j,m,new}^{DL}\phi_{m,m}\mathbf{h}_{m,l,new}^{DL}\right)^2\right) + \mathbf{B}} \tag{15}$$

$$\mathbf{B} = \sum_{b\neq j}^{B}\sum_{l\neq i}^{U_b} P_{B_b}\left((\mathbf{h}_{lb}^{DL})^2 + \sum_{m=1}^{N_{IRS}}\left(\mathbf{h}_{b,m,new}^{DL}\phi_{m,m}\mathbf{h}_{m,l,new}^{DL}\right)^2\right)$$

The achievable rates for UL and DL can then be expressed as:

$$\mathbf{c}_{ij}^{DL} = BW \times \log_2(1 + \mathbf{SINR}_{ij}^{DL}) \tag{16}$$

$$\mathbf{c}_{ijn}^{UL} = BW \times \log_2(1 + \mathbf{SINR}_{ijn}^{UL}) \tag{17}$$

### III. DELAY COMPUTATION AND UTILITY FUNCTION

This section presents the computations of delay and the definition of the utility function.

*A. Delay*

When analyzing delay, three key components are considered: 1-processing delay, 2-transmission delay and 3- queuing delay. The transmission delay for the $i$th user and $j$th AP in the nth subcarrier frequency calculated, and it represents the time needed to transmit data over the wireless link:

$$\mathbf{D}_{ijn}^T = \frac{S_i}{\mathbf{c}_{ij}^{DL}} + \frac{A_i}{\mathbf{c}_{ijn}^{UL}} \tag{18}$$

$S_i$ represents the maximum number of bits transmitted in the DL by each AP to the $i$th user. On the other hand $A_i$ denotes the size of the tracking vector in the UL, which each user transmits to the assigned AP.

The processing delay, which accounts for the time required to process the transmitted information at both the AP and user ends, is given by the following equation [10]:

$$\mathbf{D}_i^p(K_{ijn}) = \frac{v\left(l\left(X_i(\mathbf{SINR}_{ijn}^{UL})\right), l(X_i^R)\right)}{\frac{M}{N_j}} \tag{19}$$

The equation for the delay in processing information, with respect to user feedback and image transmission, is $0 \leq v\left(l\left(X_i\left(\text{SINR}_{ijn}^{UL}\right)\right), l(X_i^R)\right) \leq S_i$, where $v$ is the number of bits used to transmit the image from $l\left(X_i\left(\text{SINR}_{ijn}^{UL}\right)\right)$ to $l(X_i^R)$, calculated by each AP. $l$ is a metric based on the distance between $X_i^R$ and $X_i\left(\text{SINR}_{ijn}^{UL}\right)$. $X_i^R$ is the feedback that users send to the AP when they are dissatisfied with the displayed VR image, which is transmitted via a dedicated wireless channel. This feedback is only sent when users feel uncomfortable in the environment [10]. $M$ in (19) represents the overall processing limit of each AP. $N_j$ is the power dedicated to each user from the $j$th AP and $K_{ijn}$ signifies the accuracy in routing the $i$th user by the $j$th AP at $n$th subcarrier. Regarding the queuing delay, The distribution of user requests follows a Poisson distribution with a mean $\lambda_i$. The service time for each user's request follows an exponential distribution with a parameter $\mu_j$. The buffer size at each AP is assumed to be infinite. The inequality $\mu_j > \lambda_i$ is assumed to maintain a tradeoff between the request time of users and the service time. The queuing delay time is given by $\frac{1}{\mu_j - \lambda_i}$. The total delay time is then computed as (20),

$$\mathbf{D}_{ijn} = \mathbf{D}_i^p(K_{ijn}) + \mathbf{D}_{ijn}^T + \left(\frac{1}{\mu_j - \lambda_i}\right) \tag{20}$$

### B. Utility Function

The multi-attribute utility function in this paper takes a joint conditional form with respect to the accuracy of routing [10]. This function describes the utility between the $i$th user and the $j$th AP as $U_i(D_{ijn}, K_{ij})$ in the $n$th subcarrier. The conditional utility function, which is expressed as $U_i(D_{ijn}|K_{ijn})$, that is given by equation (21),

$$\mathbf{U}_i(\mathbf{D}_{ijn}|K_{ijn}) = \begin{cases} \dfrac{D_{max,i}(K_{ijn}) - \mathbf{D}_{ijn}}{D_{max,i}(K_{ijn}) - \gamma_{D_i}} & \mathbf{D}_{ijn} \geq \gamma_{D_i} \\ 1 & \mathbf{D}_{ijn} < \gamma_{D_i} \end{cases} \tag{21}$$

The utility function depends on the maximum tolerable delay $\gamma_{D_i}$, the $i$th user and the maximum delay $D_{max,i}(K_{ijn}) = \max_n(\mathbf{D}_{ijn})$ which is the maximum delay for the $i$th user across all subcarriers $n$. Additionally, the utility function has the following boundary conditions: $\mathbf{U}_i(D_{max,i}|K_{ijn}) = 0$, meaning that if the delay exceeds the maximum tolerable delay, the utility becomes zero, and $\mathbf{U}_i(\gamma_{D_i}|K_{ijn}) = 1$, meaning that if the delay is less than the maximum tolerable delay, the utility is maximized (equal to 1). When $\mathbf{D}_{ijn} < \gamma_{D_i}$, (within the tolerable delay), the value of the conditional utility function is equal to 1, indicating the best possible utility for the user. Thus, the total utility function is expressed as:

$$\mathbf{U}_i(\mathbf{D}_{ijn}, K_{ijn}) = \mathbf{U}_i(\mathbf{D}_{ijn}|K_{ijn})\mathbf{U}_i(K_{ijn}) \tag{22}$$

$$\mathbf{U}_i(\mathbf{D}_{ijn}, K_{ijn}) = \left(1 - \frac{\left\|X_i\left(\text{SINR}_{ijn}^{UL}\right) - X_i^R\right\|}{\max_n \left\|X_i\left(\text{SINR}_{ijn}^{UL}\right) - X_i^R\right\|}\right)\left(\frac{D_{max,i}(K_{ijn}) - \mathbf{D}_{ijn}}{D_{max,i}(K_{ijn}) - \gamma_{D_i}}\right) \tag{23}$$

The utility function $U_i(K_{ijn})$ in (22) represents the value of the utility function for the $i$th user, where $K_{ijn}$ corresponds to the accuracy in routing the user by the jth AP at the nth subcarrier. (23) is based on normalized root mean square error model. From (23) is based on the normalized root mean square error (NRMSE) model.

## IV. SOLUTION METHOD FOR OPTIMIZATION PROBLEM

The optimization problem aims to maximize the utility function, which corresponds to the maximum sum rate of users in the system. The AO method is employed to decompose each non-convex problem into simpler sub-problems. This approach is effective in handling the non-convexity typically found in wireless optimization problems, especially when beamforming and channel gain optimization are involved. By alternating between the sub-problems, the AO method ensures that the solution converges to a near-optimal state.

### A. The solution method for the optimization problem

Algorithm 2 is used for optimizing the phase shifts of the IRS elements, allowing for improved channel conditions and higher data throughput. Next, the hybrid beamforming matrices are optimized based on the computations in Section II. The optimization algorithm is solved iteratively. Each iteration refines the system's beamforming matrices and phase shift values until convergence is reached. The convergence criteria are outlined in Algorithm 2, ensuring that the system approaches an optimal solution. The rate in the DL is computed in several rounds, following the steps in Algorithm 1. This iterative process allows for the accurate computation of the DL rate as the system moves toward convergence. The total number of iterations required to solve the optimization problem is denoted as R. In the simulation, the minimum number of IRS elements is used based on [17].

### B. Phase estimation of IRS elements

The method for estimating IRS elements used in this work is the RCG method, as described in [9]. To estimate the IRS phase shift elements in the DL, Algorithm 2 is used, leveraging the rate in the DL based on the $\mathbf{c}_{ij}^{DL}$.

| **Algorithm 1** |
|---|
| (input): $N_t, N_r, N_{IRS}, user_{loc}(ul), AP_{loc}(Al), IRS_{loc}(Il)$ |
| (output): $DL_{rate}, UL_{rate}$ |
| $for\ i\ in\ ul$ |
| $for\ j\ in\ Al$ |
| $compute\ \mathbf{h}_{ijn}^{NLoS}, \mathbf{h}_{ijn}^{DL}\ (based\ on\ 3,17)$ |
| $end$ |
| $end$ |
| % UL and DL rate |
| $for\ i\ in\ ul$ |
| $for\ m\ in\ Il$ |
| $compute\ \mathbf{H}_{i,n,m}^{UL}\ for\ UL\ and\ \mathbf{h}_{m,i}^{DL}\ for\ DL\ based\ on\ (1,16)$ |
| $end$ |
| $end$ |
| $for\ m\ in\ Il$ |
| $for\ j\ in\ Al$ |
| $compute\ \mathbf{H}_{m,,j,n}^{UL}\ for\ UL\ and\ \mathbf{h}_{j,m}^{DL}\ for\ DL\ based\ on\ (2,15)$ |
| $end$ |
| $end$ |
| $compute\ \mathbf{SINR}_{ijn}^{UL}\ for\ UL\ and\ \mathbf{SINR}_{ij}^{DL}\ for\ DL\ based\ on\ (14-15), then\ compute\ rate\ in\ UL\ and\ DL\ based\ on\ (16-17)\ and\ (\textbf{Algorithm2})$ |

| **Algorithm 2** |
|---|
| (input): $initial\ \phi_{m,m}^0$ phase value, $m$ is the variable for elements of IRS. |
| (output): $\phi_{m,m}^{est}$ (estimated phases of IRS elements) |
| $compute\ \nabla \mathbf{c}_{ij}^{DL}(\phi_{m,m}^r)$ |
| $set\ \eta_m = -grad(\mathbf{c}_{ij}^{DL}(\phi_{m,m}^r))$ |
| $repeat$ |
| $choose\ Polak-Ribiere\ parameter\ \zeta^m$ |
| $update\ \eta_r^{m+1}\ by\ gradient\ ascent$ |

$$until \ (grad(\mathbf{c}_{ij}^{DL}(\phi_{m,m}^r)))_2 \leq \epsilon$$

$$Return \ \phi_{m,m}^{est}$$

### C. Complexity Analysis

In this section, we analyze the computational complexity of our proposed algorithm in comparison to methods described in the literature. The AO algorithm iteratively addresses the subproblems related to the hybrid beamforming matrices and IRS phase shift elements. We provide a closed-form solution for the hybrid beamforming matrices for the IRS phase shift elements based on Algorithm 2. The computational complexities for the IRS phase shift elements and hybrid beamforming matrices between the user-IRS and IRS-AP are as follows: $O_{IRS} = \mathcal{O}(2R \times (N_t N_r + N_t N_{IRS} N_r))$, $O_{user-IRS} = \mathcal{O}(N_{SC} \times N_r)$ and $O_{IRS-AP} = \mathcal{O}(N_{SC} \times N_t)$. For simplicity in comparison, assuming $N_t = N_r = N_{IRS} = M$. Table I summarizes the dominant computational complexities of several algorithms for computing hybrid beamforming and IRS phase shift elements in comparison to our proposed algorithm. Based on the assumption of $N_t = N_r = N_{IRS} = M$, the complexity of our proposed method is declared in MIMO channel. As shown in Table I, the computational complexity of our proposed method for estimating the hybrid beamforming matrices and phase shift elements of the IRS is lower than that of the methods in[9, 14].

TABLE I
COMPUTATIONAL COMPLEXITY COMPARISON

| | |
|---|---|
| [9] hybrid beamforming | $\mathcal{O}(3M^3 + M)$ |
| [14] hybrid beamforming | $\mathcal{O}(M^2 + M^3)$ |
| [14] IRS phase shift optimization | $\mathcal{O}(M^4)$ |
| [9] IRS phase shift optimization | $\mathcal{O}(2M^2 R N_s + M^3 + M)$ |
| Proposed hybrid beamforming | $\mathcal{O}(2M^2)$ |
| IRS phase shift optimization in proposed method | $\mathcal{O}(2R(M^2 + M^3))$ |

$$\nabla \mathbf{c}_{ij}^{DL}(\phi_{m,m}^r) = \nabla(\log_2(1 + \mathbf{SINR}_{ij}^{DL})) = \quad (24)$$
$$\frac{1}{\ln(2)} \cdot \left( \frac{\sum_{m=1}^{N_{IRS}} \left[ 2\phi_{m,m}^r \mathbf{B} - 2\phi_{m,m}^r \left( \sum_{l \neq i} P_{B_j} \left( (\mathbf{h}_{lj}^{DL})^2 + \sum_{m=1}^{N_{IRS}} (\mathbf{h}_{j,m,new}^{DL} \phi_{m,m}^r \mathbf{h}_{m,l,new}^{DL})^2 \right) \right) + \mathbf{C} \right] \mathbf{A}}{1 + \mathbf{SINR}_{ij}^{DL}} \right)$$

$$\mathbf{A} = P_{B_j} \left( (\mathbf{h}_{ij}^{DL})^2 + \sum_{m=1}^{N_{IRS}} (\mathbf{h}_{j,m,new}^{DL} \phi_{m,m}^r \mathbf{h}_{m,i,new}^{DL})^2 \right)$$

$$\mathbf{B} = \sigma^2 + \sum_{l \neq i} P_{B_j} \left( (\mathbf{h}_{lj}^{DL})^2 + \sum_{m=1}^{N_{IRS}} (\mathbf{h}_{j,m,new}^{DL} \phi_{m,m}^r \mathbf{h}_{m,l,new}^{DL})^2 \right)$$
$$+ \sum_{b \neq j}^{B} \sum_{l \neq i}^{U_b} P_{B_b} \left( (\mathbf{h}_{lb}^{DL})^2 + \sum_{m=1}^{N_{IRS}} (\mathbf{h}_{b,m,new}^{DL} \phi_{m,m}^r \mathbf{h}_{m,l,new}^{DL})^2 \right)$$

$$\mathbf{C} = 2\phi_{m,m}^r \left( \sum_{b \neq j}^{B} \sum_{l \neq i}^{U_b} P_{B_b} \left( (\mathbf{h}_{lb}^{DL})^2 + \sum_{m=1}^{N_{IRS}} (\mathbf{h}_{b,m,new}^{DL} \phi_{m,m}^r \mathbf{h}_{m,l,new}^{DL})^2 \right) \right)$$

## V. SIMULATION RESULTS

In this section, we present the performance results of the proposed algorithm through computer simulations in MATLAB and NS3. The considered environment is illustrated in Fig. 2, focusing on an indoor scenario with four users (U = 4) and two APs (B = 2). The dimensions of the indoor area are [0:10], [0:17], [0:3] in the X, Y, and Z directions, respectively.

The parameter values used in the simulations are shown in Table II. Quadrature Amplitude Modulation (QAM) with 64 subcarriers is considered in the OFDM setup. To simulate this network in NS3, the channel parameters and multipath components are first estimated in MATLAB. These parameters are then formatted in JSON and used in NS3 directories to extract SNR values as .csv files. By computing the SINR in MATLAB using these extracted SNR values from NS3, the DL and UL rates are determined. In the .csv file, each node (comprising 4 users and 2 APs) is assigned a specific ID.

Additionally, six codebooks are considered for each AP, and one type of codebook is used at each user to compare simulation results. These codebooks are generated in NS3. The NS3 code, available at the [18] as a GitHub link, is useful for this purpose. These parameters are used in below scenarios in followed six codebooks:

NS3 with no IRS, Mean channel gain SINR with no IRS, Minimum channel gain with no IRS, NS3 with 1IRS, Mean channel gain with 1IRS, Minimum channel gain with 1IRS.

We have considered six scenarios for codebooks of APs. These scenarios are as follows: "2Antenna, 1 RF chain","2Antenna, 2 RF chain","4Antenna, 1RF chain", "4Antenna, 2RF chain", "8Antenna, 1RF chain", "8Antenna, 2RF chain".

First, the simulation results of NS3 with and without IRS are compared by evaluating the mean channel gain SINR across all six codebooks of each AP, as well as the single codebook considered for each user. This comparison is illustrated in Fig. 3-5 for the first user and APs. Additionally, the DL rate in NS3 is compared with that in MATLAB using the mean channel gain in MATLAB. Both results exhibit nearly identical trends across all codebooks, as indicated on the X-axis.

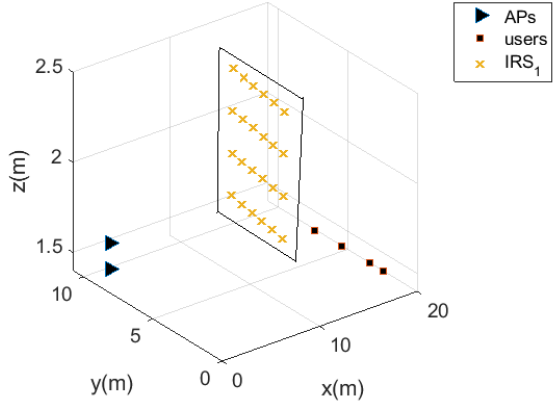

**Fig. 2.** Indoor simulation area.

Based on Fig. 3, the rate flow in NS3 when considering IRS is approximately equal for the DL rate. The DL rate in MATLAB is calculated based on the mean channel gain. The DL channel type is based on a tapped delay channel in the mmWave band, as calculated in (19), while the UL channel operates in the sub-6 GHz band and is computed according to (5).Fig. 4 illustrates the utility function values for the considered scenario across six codebooks at each AP. As seen in Fig. 4, the minimum SINR without IRS results in the lowest utility function value compared to the other scenarios, while the utility function of NS3 with IRS achieves the highest value. Fig. 5 shows the utility function values for some codebooks as a function of the number of IRS elements. As depicted in this figure, the utility function exhibits a slight decreasing trend as the number of IRS elements decreases, and the results approach those of the scenario where no IRS is used, as shown in Fig. 4. Conversely, as the number of IRS elements increases, the utility function improves, as seen in Fig. 5. This improvement occurs because, in the presence of obstacles between the user and the AP, an IRS with tunable elements can manipulate the propagation of electromagnetic waves. For clarity and to avoid confusion with scales, the X-axis labels in Fig. 5 has been adjusted to represent the number of IRS elements. Figures 6 illustrates the minimum transmission delay in the scenarios between the first user and the APs. As shown in these figures, the use of an IRS between the AP and the user generally leads to a decreasing trend in transmission delay or an incursion in DL rate across nearly all codebooks of the APs. This is due to the fact that, as shown in Fig. 6, the utility function improves when an IRS is used between the AP and the user, as demonstrated by the utility function metric. As shown in Fig. 6, the results for the minimum transmission delay based on the methods in [9, 14] are also illustrated. The results of our proposed method outperform those in [9]. This improvement is because the phase shifts of IRS elements in our proposed method are estimated using Algorithm 2 after hybrid beamforming, and then the hybrid beamforming matrices are updated following the IRS phase element estimation. This process leads to higher channel gain and, consequently, a higher DL rate. Furthermore, the channel modeling in [9] in the DL is different from that in our proposed method, which is based on equation (19). Specifically, [9] uses Rician modeling for both the AP–IRS and IRS–user channels but does not consider the NLoS path between the AP and user in the DL. In contrast, our proposed method employs OFDM modulation and incorporates the NLoS path between the AP and user based on the channel model in (19). Since [9] neither uses OFDM modulation nor accounts for the NLoS path in its channel modeling, the DL rate in their approach cannot exceed the gain of the AP–IRS–user channel alone. As shown in Fig. 6, the results for the minimum transmission delay based on the methods in [9, 14] are also illustrated. The results of our proposed method outperform those in [9]. This improvement is because the phase shifts of IRS elements in our proposed method are estimated using Algorithm 2 after hybrid beamforming, and then the hybrid beamforming matrices are updated following the IRS phase element estimation. This process leads to higher channel gain and, consequently, a higher DL rate. Furthermore, the channel modeling in [9] in the DL is different from that in our proposed method, which is based on equation (19). Specifically, [9] uses Rician modeling for both the AP–IRS and IRS–user channels but does not consider the NLoS path between the AP and user in the DL. In contrast, our proposed method employs OFDM modulation and incorporates the NLoS path between the AP and user based on the channel model in (19). Since [9] neither uses OFDM modulation nor accounts for the NLoS path in its channel modeling, the DL rate in their approach cannot exceed the gain of the AP–IRS–user channel alone.

TABLE II

PARAMETERS OF SIMULATION

| $N_{IRS}$ | 24 | $N_{SC}$ | 164 |
|---|---|---|---|
| $S_i$ | 512×24 | $v$ | 5 |
| $U$ | 4 | $\mu_j$ | 4e-9 |
| $v$ | 5 | $B$ | 2 |
| $w$ | 4.6 | $N_r$ | 1 |
| $N_{RF}$ | [1,2] | $A_i$ | 6 |
| $B$ | 2 | $\lambda_i$ | 2e-9 |

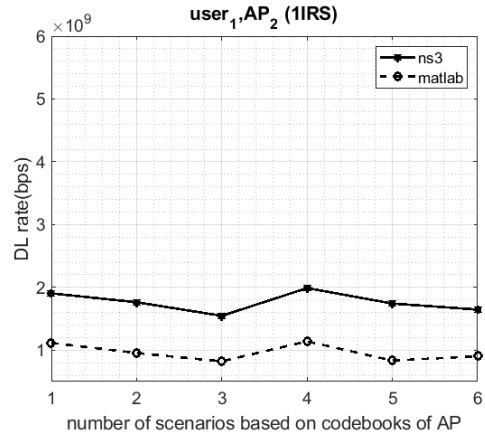

**Fig. 3.** Comparison of NS3 and MATLAB simulation of DL rate for $user_1$ and $AP_2$.

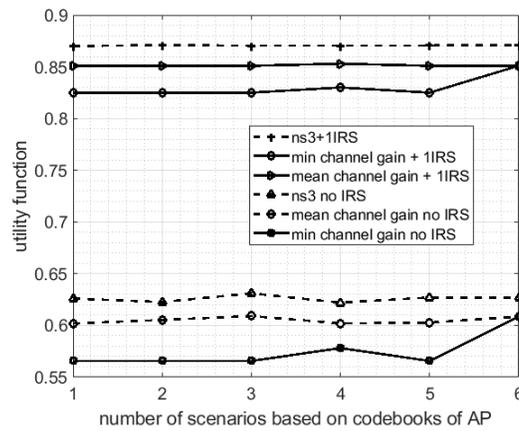

**Fig. 4.** Utility function in all scenarios and all codebooks for 24 elements of IRS.

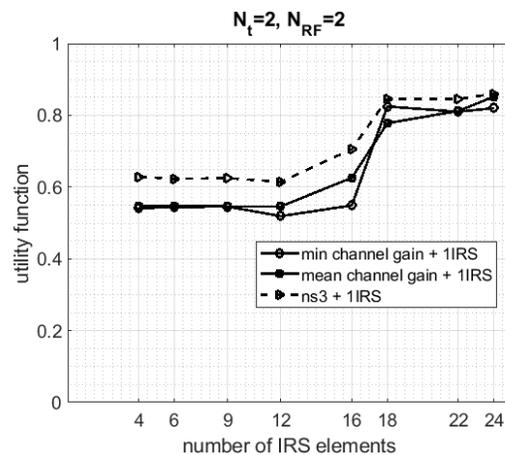

**Fig. 5.** Utility function for the second codebook in various number of elements in both of IRSs.

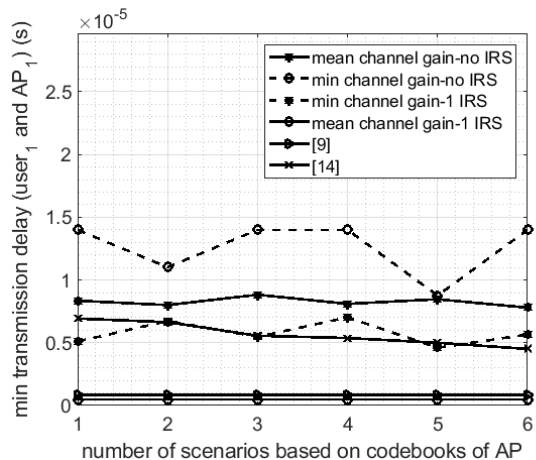

**Fig. 6** Min transmission delay between $user_1$ and $AP_1$.

In [14], the DL channel is modeled as a tapped delay channel with OFDM modulation, and the IRS phase shifts are determined by the product of the steering vectors corresponding to the IRS–user and AP–IRS channels. In contrast, in our proposed method, the IRS phase shifts are optimized through rate maximization until convergence, as described in Algorithm 2. Furthermore, unlike [14], our approach iteratively updates the IRS phase shifts after estimating the hybrid beamforming matrix, leading to enhanced channel gain and overall performance. All simulations are conducted using Ubuntu 20.04. Figure 7 illustrates $\left(grad(\nabla \mathbf{c}_{ij}^{DL}(\phi_{m,m}^r))\right)_2$ between AP1- User1 and AP2-User1 based on Algorithm 2, considering the sixth codebook defined as "8 Antenna, 2 RF chains" for APs, versus the number of iterations required for convergence. As shown in Figure 7, $\left(grad(\nabla \mathbf{c}_{ij}^{DL}(\phi_{m,m}^r))\right)_2$ decreases as the number of iterations for convergence in Algorithm 2 increases. We note that Algorithm 2 is guaranteed to converge by reaching a critical point, where the $\epsilon$ value is determined based on the DL rate. This value approaches almost zero. *f* in ylabel of Fig7 is equal to $\mathbf{c}_{ij}^{DL}$.

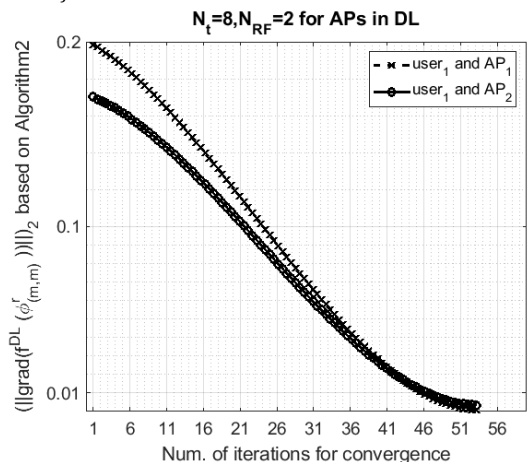

**Fig. 7.** Values of $\left(grad(\nabla \mathbf{c}_{ij}^{DL}(\phi_{m,m}^r))\right)_2$ based on Algorithm 2.

### 1.6. CONCLUSION

This paper introduces a novel multi-attribute utility function designed for VR systems, built upon the IEEE 802.11ay standard and enhanced by IRS technology. We investigate six distinct codebooks for APs and evaluate two scenarios (mean and minimum channel gain across subcarriers derived from the estimated hybrid beamforming matrices and IRS phase shifts), while comparing the utility function values and transmission delay. RCG method is used for estimation of phase shift element of IRS. Our findings indicate that increasing the number of IRS elements improves the utility function, while simultaneously decreasing transmission delay. To mitigate the costs associated with managing the phase shifts of IRS elements, we propose a strategy that minimizes the number of IRS elements by considering their maximum distance from both the user and the AP. Additionally, we compare simulation outcomes from NS3 and MATLAB, focusing on the scenario based on mean channel gain, offering valuable insights into the effectiveness of the proposed approach. With an average accuracy of 81.57%. As demonstrated in the simulation results, our proposed approach exhibits lower computational complexity and achieves a smaller minimum transmission delay in DL compared to the existing designs as mentioned in literature.